\documentclass[a4paper]{article}\usepackage[]{graphicx}\usepackage[]{color}

\usepackage[left=2.8cm, right=2.8cm, top = 2.8cm, bottom = 3cm]{geometry}
\usepackage{amsmath}
\usepackage{amssymb}
\usepackage{natbib}
\usepackage{url}

\title{An extended note on the multibin logarithmic score used in the FluSight competitions}
\author{Johannes Bracher}
\IfFileExists{upquote.sty}{\usepackage{upquote}}{}
\begin{document}

\maketitle

\begin{center}
Epidemiology, Biostatistics and Prevention Institute, University of Zurich,\\ Hirschengraben 84, 8001 Zurich, Switzerland\\

\medskip

\texttt{johannes.bracher@uzh.ch}

\bigskip

\end{center}

\begin{center}
\noindent\fbox{\parbox{0.85\textwidth}{\footnotesize
This note elaborates on a letter published in PNAS (\citealt{Bracher2019}, \url{https://doi.org/10.1073/pnas.1912147116}), which due to space restrictions contains little detail. In their reply, \citeauthor{Reich2019b}(\citeyear{Reich2019b}, \url{https://doi.org/10.1073/pnas.1912694116}) discuss the usefulness of different scoring rules in a public health context.
}}
\end{center}

\begin{abstract}
In recent years the Centers for Disease Control and Prevention (CDC) have organized \textit{FluSight} influenza forecasting competitions. To evaluate the participants' forecasts a multibin logarithmic score has been created, which is a non-standard variant of the established logarithmic score. Unlike the original log score, the multibin version is not proper and may thus encourage dishonest forecasting. We explore the practical consequences this may have, using forecasts from the 2016/17 \textit{FluSight} competition for illustration.
\end{abstract}

\section{Introduction}

In recent years the Centers for Disease Control and Prevention (CDC) have organized \textit{FluSight} influenza forecasting competitions \citep{Reich2019}, which have ``pioneered infectious disease forecasting
in a formal way'' \citep{Viboud2019}. The competitions distinguish themselves by an elaborate technical infrastructure which allows a large number of participating teams to submit weekly forecasts for several quantities in real time. All targets are based on a measure called \textit{weighted influenza-like illness} (wILI) which describes the proportion of outpatient visits due to influenza-like symptoms. Specifically, the targets are (see \citealt{Reich2019}, Fig. 1B): (a) The wILI values one to four weeks ahead of the last available observation, classified into bins of width 0.1\%. (b) The week of the season onset. (c) The peak week. (d) The peak intensity, \textit{i.e.}\ the wILI value in the peak week. All targets are discrete, and participants submit their forecasts in the form of distributions assigning a probability to each possible outcome. Forecasts are evaluated based on the multibin logarithmic score \citep{CDCP2018}, a non-standard variant of the logarithmic score \citep{Gneiting2007}. Following to its use in the \textit{FluSight} competitions this score has also been adopted in numerous scientific works (\textit{e.g.}\ \citealt{Akhmetzhanov2019}, \citealt{Ben-Nun2019}, \citealt{Brooks2018}, \citealt{Farrow2017}, \citealt{Kandula2018}, \citeyear{Kandula2019a}, \citealt{Kandula2019}, McGowan et al \citeyear{McGowan2019}, \citealt{Osthus2019a}; \citealt{Osthus2019}, \citealt{Reich2019}, \citealt{Zimmer2018}), even though it is not \textit{proper}. Propriety is generally considered an important requirement for scoring rules \citep{Gneiting2007, Held2017}, as it encourages honesty of forecasters. The goal of this note is to explore the practical implications the use of an improper scoring rule may have, detailing on a previously published letter \citep{Bracher2019}.

\section{Proper scoring rules}

The evaluation of probabilistic forecasts requires comparison of a predictive distribution $F$ for a quantity $Y$ to a single observed outcome $y_{\text{obs}}$. Various scoring rules $S(F, y_{\text{obs}})$ have been suggested to this end, and there is no single ``best'' approach. However, it is agreed that \textit{propriety} is a desirable property of a score \citep{Gneiting2007}. A score $S$ is called proper if its expectation is maximized by the true distribution of $Y$, and \textit{strictly} proper if this maximum is unique. Intuitively speaking, the highest expected score should be achieved by a forecast which is based on a perfect understanding of the system to be predicted.

Another implication of propriety is that it encourages \textit{honesty} of forecasters. To understand this, assume a forecaster whose belief about $Y$ is given by the distribution $F$. The forecaster is asked to issue a predictive distribution for $Y$, and will receive a reward depending on the agreement between this forecast and the outcome $y_{\text{obs}}$. This is measured using a score $S$, the expectation of which the forecaster consequently aims to maximize. She will do so based on her true belief $F$, but she is \textit{not} obliged to actually issue $F$ as her forecast to be evaluated. If there is a different predictive distribution $G$ so that
\begin{equation}
\mathbb{E}[S(G, Y) \mid F] > \mathbb{E}[S(F, Y) \mid F], \label{eq:G_larger_F}
\end{equation}
\textit{i.e.}\ $G$ has a higher expected score than $F$ if $F$ is actually true, the forecaster can issue $G$ instead. Such strategies are called \textit{hedging}, and ``it is generally accepted that it is undesirable to use a score for which hedging can improve the score or its expected value'' \citep[p.25]{Jolliffe2008}. To discourage hedging and the reporting of forecasts which differ from forecasters' actual beliefs, the score $S$ should be constructed such that there can be no pair of $F$ and $G$ for which \eqref{eq:G_larger_F} holds. This is exactly the definition of a proper score.

\section{The log score and the multibin log score}
\label{sec:definitions}

A widely used score is the \textit{logarithmic} or \textit{log score}. It is defined as \citep{Gneiting2007}
\begin{align*}
\text{logS}(F, y_{\text{obs}}) = \text{log}[f(y_{\text{obs}})],
\end{align*}
where $f$ is the density or probability mass function of the forecast distribution $F$. For a categorical $F$ with ordered levels $1, \dots, T$ and probabilities $p_1, \dots, p_T$ as in the \textit{FluSight} competitions this becomes
\begin{align*}
\text{logS}(F, y_{\text{obs}}) = \text{log}(p_{y_{\text{obs}}}).
\end{align*}
This score has many desirable properties \citep{Gneiting2007a}, notably it is strictly proper.

In the \textit{FluSight} competitions a modified log score is used in which not only the probability mass assigned to the observed outcome $y_{\text{obs}}$, but also the $d$ neighbouring values on either side is counted. The resulting \textit{multibin log score} \citep{CDCP2018} is defined as
\begin{align}
\text{MBlogS}(F, y_{\text{obs}}) = \text{log}\left(\sum_{i = -d}^d p_{y_{\text{obs}} + i} \right),
\end{align}
where $p_t = 0$ for $t < 1$ and $t > T$. The \textit{FluSight} organizers chose $d = 5$ for the wILI forecasts (\textit{i.e.}\ forecasts within a range of 0.5 percentage points are considered accurate) and $d = 1$ week for the onset and peak timing. The multibin log score has been argued to measure ``accuracy of practical significance'' (\citealt{Reich2019}, p.3153) while showing little sensitivity to retrospective corrections of wILI values (McGowan et al \citeyear{McGowan2019}). It has therefore been favoured over the regular log score. For both scores larger values are better and overall results are obtained by averaging over all forecasts issued by a team.

\section{A hedging strategy for the multibin log score}
\label{sec:how_to_cheat}

As also mentioned by Reich et al (\citeyear{Reich2019}), the multibin log score is improper. We now show how a forecaster can apply hedging to improve the expected score under her true belief $F$. For the following assume that $F$ assigns probability 0 to the $d$ extreme categories on either end of the support, \textit{i.e.}\
\begin{equation}
p_1 = \dots = p_d = p_{T - d + 1} = \dots = p_T = 0,\label{eq:regularity_condition}
\end{equation}
where $T > 2d$. This rids us of extra considerations on these categories and can always be achieved by adding categories to the support. Then denote by $\tilde{F}$ a distribution with the same support as $F$ and
\begin{align}
\tilde{p}_{t} = \frac{\sum_{i = -d}^d p_{t + i}}{2d + 1}, t = 1, \dots, T,
\label{eq:F_tilde}
\end{align}
where again $p_t = 0$ for $t < 1$ and $t > T$. This represents a ``blurred'' version of $F$, where we always re-distribute the probability mass for one outcome equally between itself and the $d$ neighbouring ones on either side (\textit{i.e.}\ the $\tilde{p}_t$ are ``moving averages'' of the $p_t$). Condition \eqref{eq:regularity_condition} ensures that $\sum_{t = 1}^T \tilde{p}_t = 1$ so that $\tilde{F}$ is a well-defined distribution. The multibin log score of $F$ can now be expressed through the regular log score of $\tilde{F}$, as
\begin{align*}
\text{MBlogS}(F, y_{\text{obs}}) = \text{logS}(\tilde{F}, y_{\text{obs}}) + \text{log}(2d + 1). \label{eq:equiv}
\end{align*}
Applying the MBlogS is thus essentially the same as applying the regular log score, but after ``blurring'' the predictive distribution as in equation \eqref{eq:F_tilde}. As the regular log score is proper, a forecaster then has an incentive to issue a sharper forecast $G$ so that the corresponding blurred distribution $\tilde{G}$ (with probabilities $\tilde{p}_{G, 1}, \dots, \tilde{p}_{G, T}$ derived from $p_{1}, \dots, p_{ T}$ in analogy to \eqref{eq:F_tilde}) is as close as possible to $F$. If there is a $G$ so that $\tilde{G}$ corresponds exactly to $F$ it can be found via recursive computations. Otherwise an optimal $G$ is obtained by numerically maximizing
$\sum_{t = 1}^T p_t \cdot \text{log}(\tilde{p}_{G, t})$
with respect to $p_{G, 1}, \dots, p_{G, T} \in [0, 1]$ while $\sum_{t = 1}^T p_{G, t} = 1$. This is the same as minimizing the Kullback-Leibler divergence \citep{Joyce2011} of $F$ and $\tilde{G}$. The optimum is not necessarily unique, but in general $G \neq F$ holds, and $G$ implies less variability than $F$.

\begin{figure*}[h!]
\center
\includegraphics[width=\textwidth]{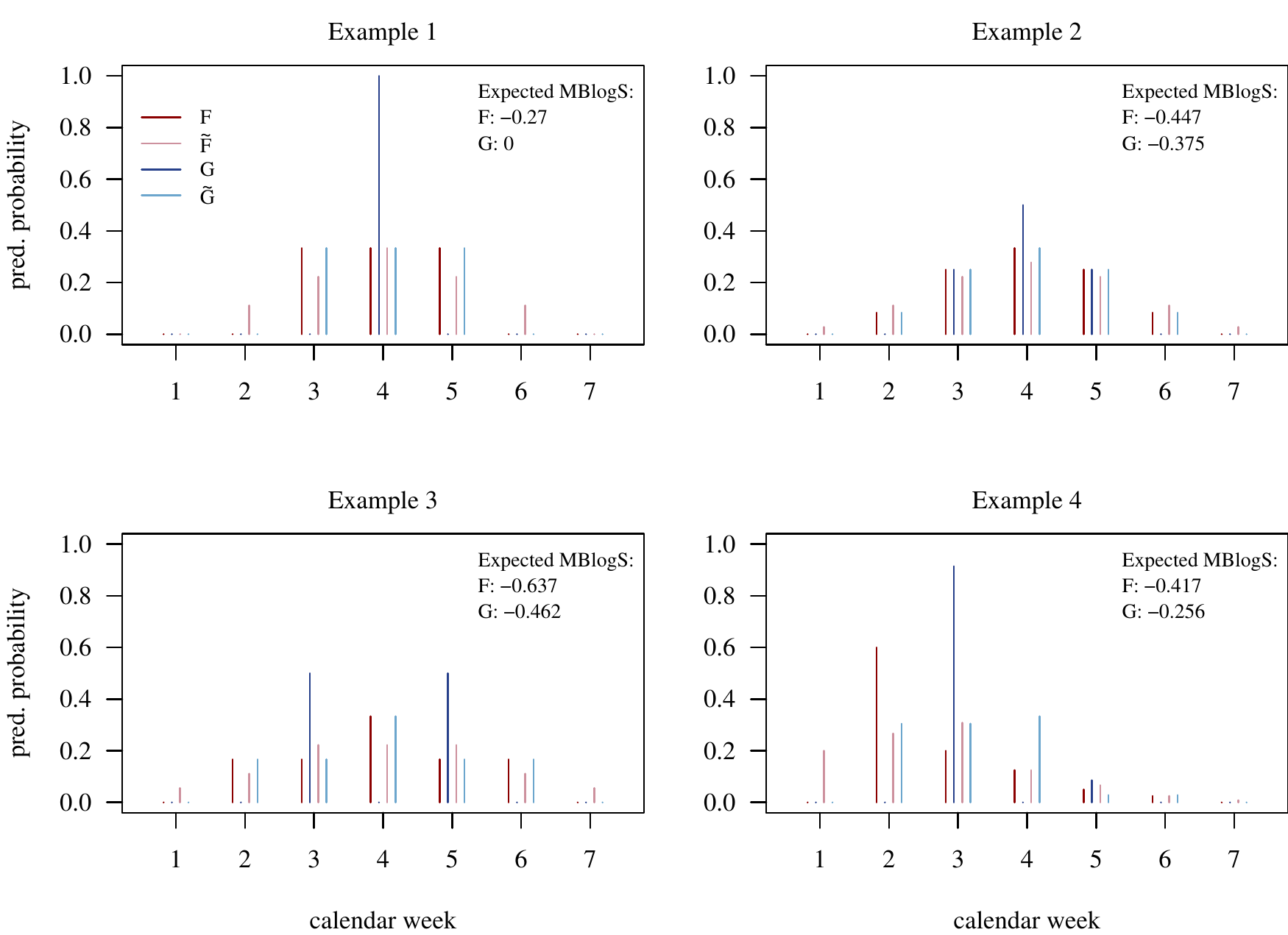} 
\caption{Examples 1--4: Original forecasts $F$, optimized versions $G$ and the respective blurred distributions $\tilde{F}$ and $\tilde{G}$. Note that $F$ and $\tilde{G}$ are identical in Examples 1--3, but not 4. Expected scores are computed under $F$.}
\label{fig:example}
\end{figure*}

Intuitively speaking, a forecaster is incentivized to issue a sharper, more ``risky'' forecast because the MBlogS does not sanction a low probability assigned to the observed value $y_{\text{obs}}$ as long as the neighbouring weeks or bins received enough probability mass. Indeed, the optimized forecast $G$ will often cover outcomes with a high probability under $F$ exclusively by assigning probability mass to their neighbours. We illustrate this using some example forecasts of the peak timing (\textit{i.e.}\ $d = 1$), visualized in Figure \ref{fig:example}:

\begin{description}
\item \textit{Example 1:} Assume we are sure that the peak of the season will occur between weeks 3 and 5, more precisely $F$ is given by $p_3 = p_4 = p_5 = 1/3$. The expected MBlogS under $F$ when reporting $F$ is $1/3 \cdot \log(2/3) + 1/3 \cdot \log(1) + 1/3 \cdot \log(2/3) = -0.270$. If, however, we report $G$ with $p_{3} = 0, p_4 = 1, p_5 = 0$, \textit{i.e.}\ claim to be sure that the peak occurs in week 4, we can expect a score of $1/3 \cdot \log(1) + 1/3 \cdot \log(1) + 1/3 \cdot \log(1) = 0$. In fact, $G$ will score at least as good as $F$ for all three outcomes we consider possible, and better for two of them.
\item \textit{Example 2:} Our true belief $F$ is given by $p_1 = 0, p_2 = 1/12, p_3 = 1/4, p_4 = 1/3, p_5 = 1/4, p_6 = 1/12, p_7 = 0$. We thus believe that the peak will occur around week 4, but even weeks 2 and 6 are considered possible. In this case we can find $G$ so that $\tilde{G}$ corresponds exactly to $F$. $G$ is given by $p_{G, 1} = p_{G, 2} = 0, p_{G, 3} = 1/4, p_{G, 4} = 1/2, p_{G, 5} = 1/4, p_{G, 6} = p_{G, 7} = 0$. We should thus claim the peak to definitely occur in weeks 3, 4 or 5. The corresponding expectations for the multibin score under $F$ are $\mathbb{E}[\text{MBlogS}(F, Y) \mid F] = -0.447$ and $\mathbb{E}[\text{MBlogS}(G, Y) \mid F] = -0.375$.
\item \textit{Example 3:} Now assume $F$ to be $p_1 = 0, p_2 = 1/6, p_3 = 1/6, p_4 = 1/3, p_5 = 1/6, p_6 = 1/6, p_7 = 0$, which is similar to the previous example, but with more probability assigned to weeks 2 and 6. Now $G$ is given by $p_1 = p_2 = 0, p_3 = 0.5, p_4 = 0, p_5 = 0.5, p_6 = p_7 = 0$. The optimized forecast distribution is thus bimodal and the peak is claimed to occur either in week 3 or 5. The expected scores are $\mathbb{E}[\text{MBlogS}(F, Y) \mid F] = -0.637$ and $\mathbb{E}[\text{MBlogS}(G, Y) \mid F] = -0.462$.
\item \textit{Example 4:} Lastly assume $F$ to be $p_1 = 0, p_2 = 0.6, p_3 = 0.2, p_4 = 0.125, p_5 = 0.05, p_6 = 0.025, p_7 = 0$. We thus consider it likely that the peak occurs in week 2, but it may also occur later. In this setting there is no $G$ so that $\tilde{G}$ corresponds exactly to $F$, but numerical optimization returns $p_1 = p_2 = 0, p_3 \approx 0.91, p_4 = 0, p_5 \approx 0.09, p_6 = p_7 = 0$. To get the highest expected score we should thus shift the mode of our predictive distribution and claim to be almost certain that the peak occurs in week 3. The expected scores are $\mathbb{E}[\text{MBlogS}(F, Y) \mid F] = -0.417$ and $\mathbb{E}[\text{MBlogS}(G, Y) \mid F] = -0.256$.
\end{description}

\noindent The patterns observed here also occur in many other settings. The optimized forecasts $G$ are sharper than the respective $F$. Moreover, the mode often gets shifted by up to $d$ weeks, and one or several additional local modes can occur.

\section{Application to \textit{FluSight} forecasts}
\label{sec:application}

To stress its practical relevance we apply the hedging strategy from the previous section to real forecasts from the \textit{FluSight} competitions. These are publicly available at \url{https://github.com/FluSightNetwork/cdc-flusight-ensemble/}. Specifically we consider national level forecasts from the 2016/17 season submitted by the Los Alamos National Laboratories (LANL) team. Their dynamic Bayesian forecasting method has shown remarkably good performance over several years \citep{Osthus2019}. We follow the same evaluation procedure as in \cite{Reich2019}, where average scores are only computed from the relevant parts of the season (\textit{e.g.}\ forecasts of the onset week are ignored once it is clear that the onset has occurred; see p.8 in \citealt{Reich2019}).

\begin{figure}[h!]
\includegraphics[width=\textwidth]{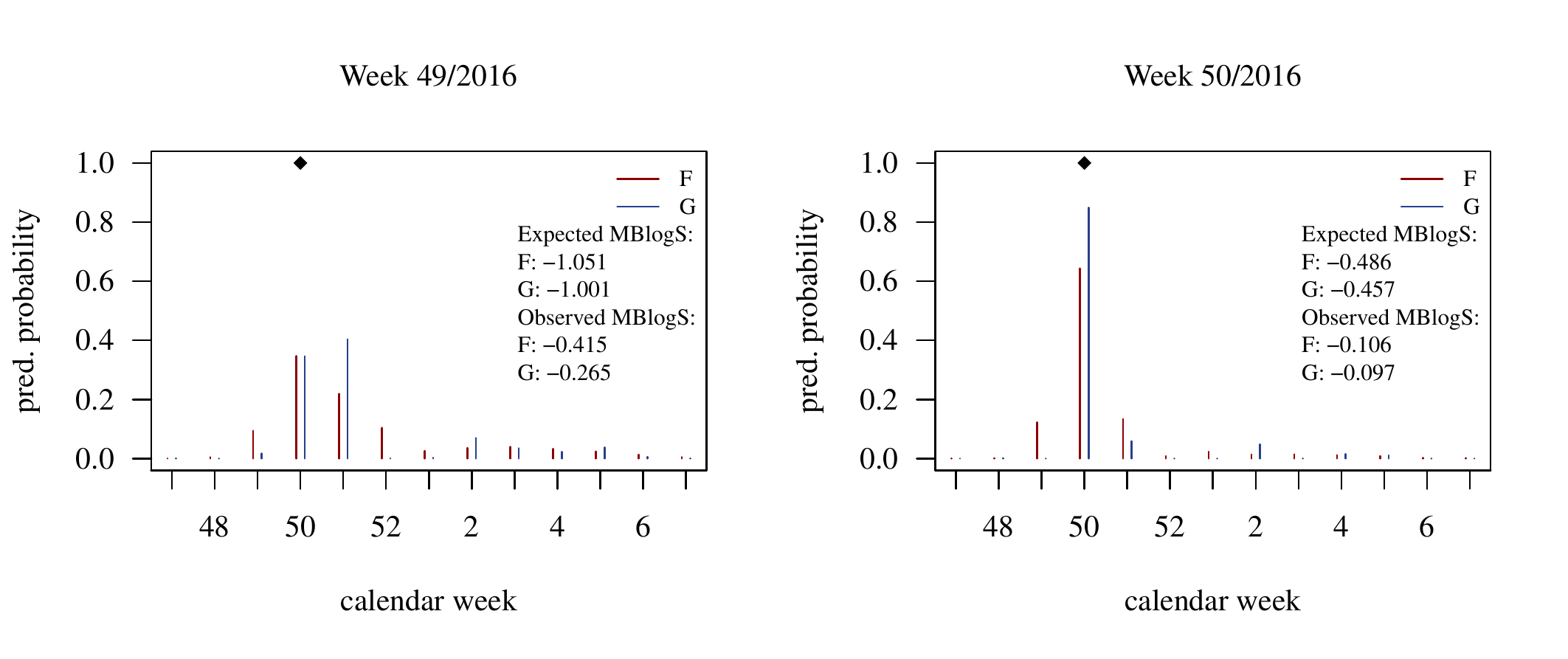} 
\caption{Forecasts $F$ for the onset week, submitted by the LANL team in weeks 49--50, 2016, and optimized versions $G$. Diamonds mark the true peak week. Expected scores are computed under $F$.}
\label{fig:osthus_detailed}
\end{figure}

For all forecasts of the seven targets (one to four-week-ahead wILI, onset and peak timing, peak incidence) we obtained optimized versions with the respective value of $d$. For illustration Figure \ref{fig:osthus_detailed} shows forecasts of the onset timing issued in calendar weeks 49 and 50, 2016. As in Example 4, the optimized forecast $G$ in week 49 has its mode shifted by one week. In both cases the optimized $G$ are visibly sharper and are multimodal, even though the corresponding $F$ are unimodal. Averaged over the 2016/17 season the optimized forecasts yield indeed higher and thus improved MBlogS for the onset timing ($-0.33$ vs. $-0.39$).

\begin{figure}[h!]
\includegraphics[width=\textwidth]{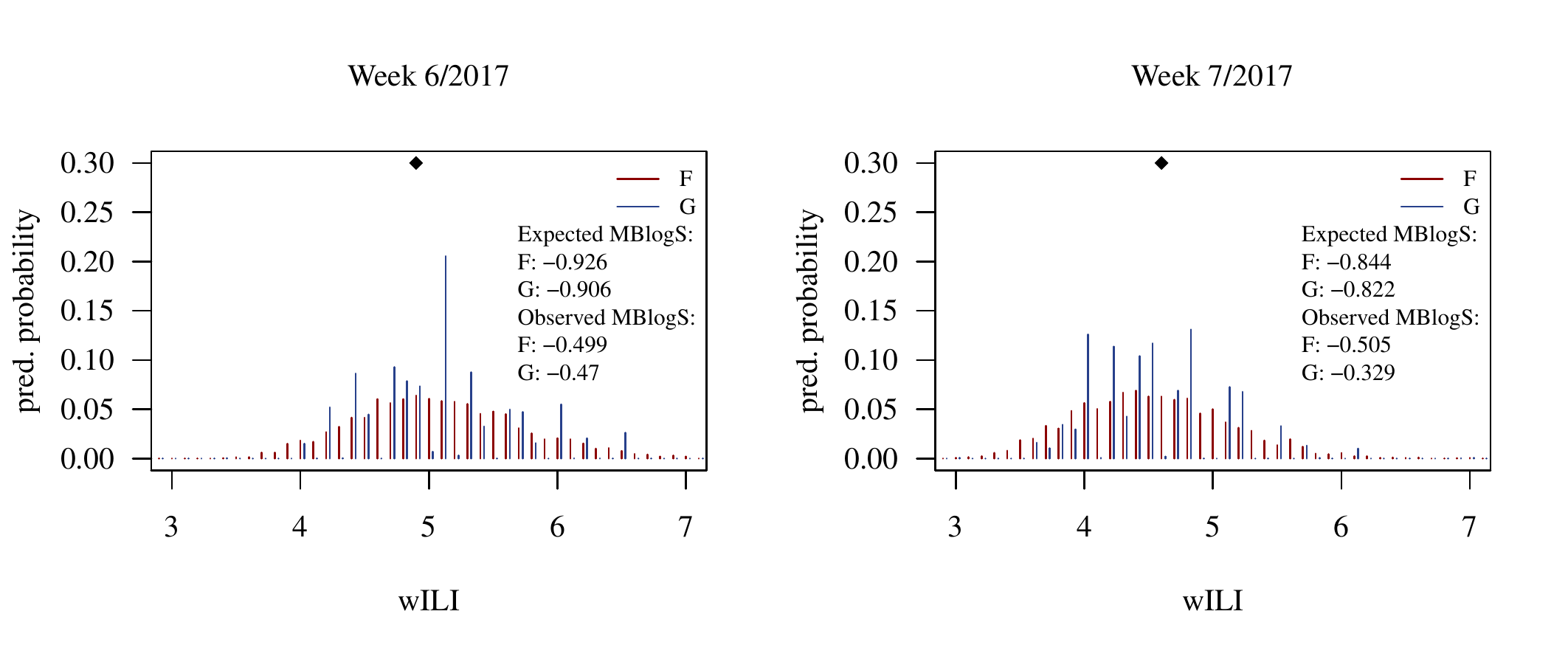} 
\caption{Forecasts $F$ for wILI (one week ahead), submitted by the LANL team in weeks 6--7, 2017, and optimized versions $G$. Diamonds mark the observed wILI values. Expected scores are computed under $F$.}
\label{fig:wILI1}
\end{figure}

Figure \ref{fig:wILI1} shows the same for one-week-ahead forecasts of wILI, \textit{i.e.}\ now we use $d = 5$. The optimized $G$ leave gaps between the values to which they assign positive probabilities. These forecast distributions with many spikes are unlikely to be useful to public health experts. Nonetheless, averaged over the course of the season, the optimized forecasts outperform the original ones ($-0.19$ vs. $-0.30$). Indeed, as shown in Table \ref{tab:mbls}, such improvements are also achieved for the remaining five targets. This illustrates that the hedging strategy enabled by the improper MBlogS can lead to non-negligible improvements of average scores in practice.

\begin{table}[h!]
\caption{Average multibin log scores for different targets at the national level over the course of the 2016/17 season: original forecasts $F$ as issued by the LANL team and optimized forecasts $G$ (with $d = 1$ for onset and peak timing, $d = 5$ otherwise).}
\begin{center}
\begin{tabular}{r r r r r r r r}
\hline
 & 1 wk & 2 wk & 3 wk & 4 wk & onset week & peak week & peak intensity \\ 
  \hline
original forecasts & -0.30 & -0.81 & -0.85 & -0.89 & -0.39 & -0.48 & -0.62 \\ 
  optimized forecasts & -0.19 & -0.75 & -0.78 & -0.84 & -0.33 & -0.43 & -0.59 \\ 
   \hline

\end{tabular}
\label{tab:mbls}
\end{center}
{\footnotesize \noindent Note that the numbers given here are not directly comparable to the ones in \cite{Reich2019b}, Fig. 1. We focus on the season 2016/17 and the national level, while \citeauthor{Reich2019b} average over different seasons and geographical resolutions.}
\end{table}

\section{Discussion}
\label{sec:discussion}

We showed that the multibin log score used in the CDC \textit{FluSight} competitions incentivizes hedging, \textit{i.e.}\ tuning forecast distributions in a specific way before submission. While we strongly doubt that participants have consciously tried to game the score, it is possible that this happens unintentionally. In forecasting, cross-validation methods to optimize forecasts for a given evaluation metric are common. Such methods could lead to hedging of the score without the authors being aware. As in previous work \citep{Held2017} we therefore advocate the use of proper scoring rules to evaluate epidemic forecasts, \textit{e.g.}\ the regular log score. Measures which are easier to interpret could be reported as a supplement to facilitate communication with public health experts.

\medskip

{ \noindent
\textbf{Data and code availability:} The data used in this note are available from the \textit{FluSight Collaboration} at \url{https://github.com/FluSightNetwork/cdc-flusight-ensemble/}. \texttt{R} codes to reproduce the presented results are available at \url{https://github.com/jbracher/multibin}.}

\medskip

{ \noindent
\noindent \textbf{Acknowledgements:} I would like to thank T. Gneiting for helpful discussions and the \textit{FluSight Collaboration} for making its forecasts publicly available. I also thank N.G. Reich for a very interesting discussion on the points raised in my letter.}

{
\footnotesize
\setlength{\bibsep}{0.5pt plus 0.3ex}
\bibliographystyle{apalike}
\bibliography{bib_mbls}

\begin{thebibliography}{}

\bibitem[Akhmetzhanov et~al., 2019]{Akhmetzhanov2019}
Akhmetzhanov, A.~R., Lee, H., Jung, S., Kayano, T., Yuan, B., and Nishiura, H.
  (2019).
\newblock Analyzing and forecasting the {E}bola incidence in {N}orth {K}ivu,
  the {D}emocratic {R}epublic of the {C}ongo from 2018--19 in real time.
\newblock {\em Epidemics}, 27:123 -- 131.

\bibitem[Ben-Nun et~al., 2019]{Ben-Nun2019}
Ben-Nun, M., Riley, P., Turtle, J., Bacon, D.~P., and Riley, S. (2019).
\newblock Forecasting national and regional influenza-like illness for the
  {USA}.
\newblock {\em PLOS Computational Biology}, 15(5):1--20.

\bibitem[Bracher, 2019]{Bracher2019}
Bracher, J. (2019).
\newblock On the multibin logarithmic score used in the {FluSight}
  competitions.
\newblock {\em Proceedings of the National Academy of Sciences}, in press, Sep
  2019, \texttt{https://doi.org/10.1073/pnas.1912147116}.

\bibitem[Brooks et~al., 2018]{Brooks2018}
Brooks, L.~C., Farrow, D.~C., Hyun, S., Tibshirani, R.~J., and Rosenfeld, R.
  (2018).
\newblock Nonmechanistic forecasts of seasonal influenza with iterative
  one-week-ahead distributions.
\newblock {\em PLOS Computational Biology}, 14(6):1--29.

\bibitem[{Centers for Disease Control and Prevention}, 2018]{CDCP2018}
{Centers for Disease Control and Prevention} (2018).
\newblock {Preliminary Guidelines for the 2018-19 Influenza Forecasting
  Challenge}.
\newblock Accessible online at
  \url{https://predict.cdc.gov/api/v1/attachments/flusight%202018%E2%80%932019/flu_challenge_2018-19_tentativefinal_9.18.18.docx},
  retrieved on 23 April 2019.

\bibitem[Farrow et~al., 2017]{Farrow2017}
Farrow, D.~C., Brooks, L.~C., Hyun, S., Tibshirani, R.~J., Burke, D.~S., and
  Rosenfeld, R. (2017).
\newblock A human judgment approach to epidemiological forecasting.
\newblock {\em PLOS Computational Biology}, 13(3):1--19.

\bibitem[Gneiting et~al., 2007]{Gneiting2007a}
Gneiting, T., Balabdaoui, F., and Raftery, A.~E. (2007).
\newblock Probabilistic forecasts, calibration and sharpness.
\newblock {\em Journal of the Royal Statistical Society: Series B (Statistical
  Methodology)}, 69(2):243--268.

\bibitem[Gneiting and Raftery, 2007]{Gneiting2007}
Gneiting, T. and Raftery, A.~E. (2007).
\newblock Strictly proper scoring rules, prediction, and estimation.
\newblock {\em Journal of the American Statistical Association},
  102(477):359--378.

\bibitem[Held et~al., 2017]{Held2017}
Held, L., Meyer, S., and Bracher, J. (2017).
\newblock Probabilistic forecasting in infectious disease epidemiology: the
  13th {A}rmitage lecture.
\newblock {\em Statistics in Medicine}, 36(22):3443--3460.

\bibitem[Jolliffe, 2008]{Jolliffe2008}
Jolliffe, I.~T. (2008).
\newblock The impenetrable hedge: a note on propriety, equitability and
  consistency.
\newblock {\em Meteorological Applications}, 15(1):25--29.

\bibitem[Joyce, 2011]{Joyce2011}
Joyce, J. (2011).
\newblock Kullback-leibler divergence.
\newblock In Lovric, M., editor, {\em International {E}ncyclopedia of
  {S}tatistical {S}cience}, pages 720--722. Springer, Berlin.

\bibitem[Kandula et~al., 2019]{Kandula2019a}
Kandula, S., Pei, S., and Shaman, J. (2019).
\newblock Improved forecasts of influenza-associated hospitalization rates with
  {G}oogle search trends.
\newblock {\em Journal of The Royal Society Interface}, 16(155):20190080.

\bibitem[Kandula and Shaman, 2019]{Kandula2019}
Kandula, S. and Shaman, J. (2019).
\newblock Near-term forecasts of influenza-like illness: An evaluation of
  autoregressive time series approaches.
\newblock {\em Epidemics}, 27:41--51.

\bibitem[Kandula et~al., 2018]{Kandula2018}
Kandula, S., Yamana, T., Pei, S., Yang, W., Morita, H., and Shaman, J. (2018).
\newblock Evaluation of mechanistic and statistical methods in forecasting
  influenza-like illness.
\newblock {\em Journal of The Royal Society Interface}, 15(144):20180174.

\bibitem[McGowan et~al., 2019]{McGowan2019}
McGowan, C., Biggerstaff, M., Johansson, M., Apfeldorf, K., Ben-Nun, M.,
  Brooks, L., Convertino, M., Erraguntla, M., Farrow, D., Freeze, J., Ghosh,
  S., Hyun, S., Kandula, S., Lega, J., Liu, Y., Michaud, N., Morita, H., Niemi,
  J., Ramakrishnan, N., Ray, E., Reich, N., Riley, P., Shaman, J., Tibshirani,
  R., Vespignani, A., Zhang, Q., Reed, C., and {The Influenza Forecasting
  Working Group} (2019).
\newblock Collaborative efforts to forecast seasonal influenza in the {U}nited
  {S}tates, 2015--2016.
\newblock {\em Scientific Reports}, Article Nr. 683.

\bibitem[Osthus et~al., 2019a]{Osthus2019a}
Osthus, D., Daughton, A.~R., and Priedhorsky, R. (2019a).
\newblock Even a good influenza forecasting model can benefit from
  internet-based nowcasts, but those benefits are limited.
\newblock {\em PLOS Computational Biology}, 15(2):1--19.

\bibitem[Osthus et~al., 2019b]{Osthus2019}
Osthus, D., Gattiker, J., Priedhorsky, R., and Del~Valle, S.~Y. (2019b).
\newblock Dynamic {B}ayesian influenza forecasting in the {U}nited {S}tates
  with hierarchical discrepancy (with discussion).
\newblock {\em Bayesian Analysis}, 14(1):261--312.

\bibitem[Reich et~al., 2019a]{Reich2019b}
Reich, N., Osthus, D., Ray, E., Yamana, T., Biggerstaff, M., Johansson, M.,
  Rosenfeld, R., and Shaman, J. (2019a).
\newblock Scoring probabilistic forecasts to maximize public health
  interpretability.
\newblock {\em Proceedings of the National Academy of Sciences}, in press, Sep
  2019, \texttt{https://doi.org/10.1073/pnas.1912694116}.

\bibitem[Reich et~al., 2019b]{Reich2019}
Reich, N.~G., Brooks, L.~C., Fox, S.~J., Kandula, S., McGowan, C.~J., Moore,
  E., Osthus, D., Ray, E.~L., Tushar, A., Yamana, T.~K., Biggerstaff, M.,
  Johansson, M.~A., Rosenfeld, R., and Shaman, J. (2019b).
\newblock A collaborative multiyear, multimodel assessment of seasonal
  influenza forecasting in the {U}nited {S}tates.
\newblock {\em Proceedings of the National Academy of Sciences},
  116(8):3146--3154.

\bibitem[Viboud and Vespignani, 2019]{Viboud2019}
Viboud, C. and Vespignani, A. (2019).
\newblock The future of influenza forecasts.
\newblock {\em Proceedings of the National Academy of Sciences},
  116(8):2802--2804.

\bibitem[Zimmer et~al., 2018]{Zimmer2018}
Zimmer, C., Leuba, S.~I., Yaesoubi, R., and Cohen, T. (2018).
\newblock Use of daily internet search query data improves real-time
  projections of influenza epidemics.
\newblock {\em Journal of The Royal Society Interface}, 15(147):20180220.

\end{thebibliography}
}

\end{document}